\documentstyle[11pt]{article}

\begin{document}

\title{
\vspace{1.0in}
No-Lose `Theorem' for Parity Violating \\
Nucleon-Nucleon Scattering Experiments\footnote {Talk given at the BNL
workshop on future directions in particle and nuclear physics at
multi-GeV \\
hadron beam facilities, March 4--6, 1993.}
\\}

\author{ T.\ Goldman \\
Los Alamos National Laboratory \\
Los Alamos, NM 87545 \\ }
\date{}

\maketitle

\vspace{-3.0in}
\begin{flushright}
LA-UR-93-1532\\
nucl-th/9304020
\end{flushright}
\vspace{3.5in}

\begin{abstract}

A purely left-chiral model of the weak interactions is used to show
that the total parity-violating asymmetry in quark-quark scattering
must grow with increasing energy. In the absence of other new physics,
non-observation of a large asymmetry can therefore be used to infer an
upper bound on the mass scale for new right-chiral weak vector bosons.
Applying this idea to actual nucleon-nucleon scattering requires more
involved calculations, as the dominant contribution appears to come from
a component of diquark-quark scattering related to, but not identical to,
wavefunction-mixing. Earlier criticism of this model by Simonius and
Unger is refuted and a new calculation is proposed as an additional
check on the result.  Finally, we argue that the so-called `spin
crisis' does not affect our conclusions.

\end{abstract}

\vspace{1.5in}

\pagebreak

\section{Hadronic Scattering in a Purely Left-Handed World}

Consider a world in which the weak interaction is described by a purely
left-chiral SU(2) theory, spontaneously broken in the usual way. To
one-loop order, there are then finite weak corrections to the
left-chiral quark-{\bf gluon} vertex function which do not exist for
the corresponding right-chiral vertex. These (finite after wavefunction
renormalization) vertex corrections have the structure of form-factors
and so imply that the left-chiral quark-gluon form-factor falls more
rapidly with increasing (spacelike) squared four-momentum of the gluon
than does the right-chiral one.

The right-chiral quark-gluon vertex is completely unaffected in this
regard because the right-chiral quarks do not couple at all to this
weak interaction.  The difference of these vertex strengths then
induces a parity violation when, say, a beam of right-chiral quarks
impinging on a target of left-chiral quark is compared with the same
beam impinging on a target of right-chiral quarks. The strong
interactions (gluon exchanges) are equal, in the two cases, as are the
weak interactions at tree level, but the latter differ at one-loop
because of the vertex correction to the left-chiral quark-gluon
coupling. Note that in this simplified model, there are no other
one-loop graphs (box or crossed-box), as in the actual case, again
because the right-chiral quarks do not participate at all in this weak
interaction.

Thus, the gluon-mediated scattering of left-chiral quarks on
right-chiral quarks falls off more rapidly with four-momentum transfer
than does the scattering of right-chiral quarks on right-chiral
quarks.  This parity-violating asymmetry (PVA) in quark-quark
scattering, defined here as the difference between the right-chiral on
right-chiral total cross-section and the left-chiral on right-chiral
one divided by their sum, develops and grows as larger and larger
four-momentum transfers contribute to the total cross-sections as the
phase space expands with increasing energy.  The statement holds for
`elastic' quark-quark scattering below the weak vector boson production
threshold.

Above that threshold, the opening channels restore total cross-section
strength and damp out the PVA because, even in this simplified model,
bremsstrahlung of weak-bosons will compensate for the form-factor
reduction of elastic quark-quark scattering. However, if the mass of
the weak vector boson is large, it should be feasible, in principle,
(unlike the corresponding electromagnetic case) to experimentally
separate out the (weak-)elastic events from all others.

The below threshold growth of the PVA can only be ameliorated by the
slow, logarithmic change with scale of the coupling constants, unless
there are new right-chiral weak interactions which appear at some
(higher) mass scale. These new interactions weaken the right-chiral
quark-gluon vertex in a corresponding fashion, stopping the growth in
the difference between the two cross-sections. As the average total
cross-section continues to grow, the fractional PVA decreases. Thus we
expect two possibilities for PVA measurements at medium and high
energies: Either the Standard Model is correct and the PVA increases
with energy,  becoming less difficult to measure; or the PVA decreases
and experiments can only put an upper bound on its value at higher
energies. In the latter case, however, we can infer an {\bf upper}
bound on the mass scale of the vector bosons for a new right-chiral
weak interaction. This is the no-lose `theorem' referred to in the
title.

Our explicit calculations~\cite{jerry} of the vertex function in the
pure SU(2)$_{L}$ case show decrements in the vertex strength of only a
tiny fraction of a percent of the dimensionless weak coupling constant
at four-momentum transfers of order 1~GeV, but this grows by more than
an order of magnitude by 10~GeV. It grows to 30\% by the TeV region.
Thus, the only practical questions are whether the PVA can be measured
at all at lower energies, and whether the theoretical calculation of
the effect in the Standard Model is consistent with those results. We
believe the answer is positive in both regards~\cite{dean}, but this
contention relies heavily on the results from an
experiment~\cite{lockyer} at ANL at 6 GeV/$c$ and a more recent
result~\cite{vince} from LAMPF at even lower energies.

\section{A Brief Digression on Low-Energy Measurements and Theories}

Several theoretical and experimental attempts have been made in the
past to measure and predict the PVA in nucleon-nucleon scattering, now
defined as the difference between the total cross-sections for positive
and negative helicity (right- and left-chiral) nucleons on unpolarized
targets divided by their sum.  Early low energy experiments were
performed at energies of 15 and 45 MeV, finding negative PVA's of order
$10^{-7}$, to no one's great surprise. The theoretical efforts~\cite{bunch}
naturally approach the problem from the point of view of
meson-exchange forces and do produce the correct sign and order of
magnitude. The main problem was to obtain independent estimates of the
parity-violating meson-nucleon vertex function strengths. Some are
available through current algebra and bounds may be obtained from
limits on parity-violating admixtures in nuclear states.  A brave
attempt was made to calculate these quantities using QCD~\cite{ddh}.
Very recently, low energy polarized neutron scattering at LAMPF has
provided a new way to extract this information at zero
energy~\cite{dave}. These experiments are especially interesting as
parity-violating effects above the 10\% level have been discovered,
compromising the old arguments that such effects must always be very
small.

However, a common feature of the calculations emerged as they were
extended to intermediate energies:  As more partial waves or channels
($\Delta$ intermediate states, and multiple meson exchanges for
example) are added, the PVA tends to become positive (the cross-over
energy has been the subject of experimental efforts at
TRIUMF~\cite{vanO}) and an envelope of the collection of curves
appears, which continues growing with energy even though each new
individual contribution eventually falls off. For recent sophisticated
efforts that include references to earlier work, see~\cite{silbar}. The
impetus for these extensions was the `high' energy result at ANL, which
seemed incomprehensibly large, at $2 \times 10^{-6}$, compared to the
model results, despite their clear inadequacy at such a relatively
large energy.

The envelope feature, however, was reminiscent to me of features I had
seen in other attempts to describe high energy phenomena in terms of
mesons and baryons, phenomena which were more readily explained (i.e.,
in terms of fewer independent amplitudes) by using quarks and gluons.
At the instigation of Darragh Nagle, I therefore set out  with Dean
Preston, to calculate the relevant components of the weak Hamiltonian
for quarks in detail, and to apply it to nucleon-nucleon
scattering~\cite{dean,me}.

\section{The QCD Calculation}

Returning now to the real world, we find many dilution factors for the
PVA signal: Due to confinement, one must use proton beams and nucleon
targets, not quarks, with the attendant averaging over different
chirality quark beams and targets. Of course, since the quarks are not
directly visible, for a differential cross-section, a high $Q^2$ jet
would have to be observed to ascertain that a large momentum transfer
did indeed occur in the scattering event. Box and crossed-box graphs
appear. The mixing of the U(1) factor with both chiralities of
couplings to include electromagnetism in the standard electroweak model
also produces some cancelling right-chiral interaction corrections.
However, it is at least very unlikely (and was indeed found not to be
the case in the restricted energy range calculations of Ref.~\cite{me})
for there to be a strong cancellation in both proton-proton and
proton-neutron scattering, (where the beam proton is polarized).

{}From a high energy, quark point of view, the strong amplitude, A, is
schematically
$$
A \sim \alpha_s/q^2 \; , \eqno (1)
$$
where $\alpha_s$ is the strong coupling, and $q^2$ is the four-momentum
transfer, and the weak parity-violating amplitude, B, is similarly
$$
B \sim G_F  \; . \eqno (2)
$$
Then, since $\alpha_s \sim 1$ and $< q^2> \sim$ 0.1 GeV$^2/c^2$,
$$
PVA \equiv \frac{\sigma_{+} - \sigma_{-}}{\sigma_{+} + \sigma_{-}}
$$
$$
\sim \frac{|A^*B|}{|A|^2}
$$
$$
\sim G_F<q^2>/\alpha_s
$$
$$
\sim 10^{-6} \; , \eqno (3)
$$
where $\sigma_{\pm}$ are the total cross sections for positive and
negative helicity nucleons on unpolarized targets. Note the general
feature that the PVA is given by the overlap of a strong and weak
amplitude, divided by the total strong interaction cross-section.

The result in Eqn.~(3) suggests the ANL result is of normal size.
However, when averaged over realistic quark distributions in nucleons
and when all the 2's and $\pi$'s are included, this result is reduced
by about one order of magnitude. More detailed calculations require
studying strong and weak amplitudes involving box and crossed-box
graphs of gluons and weak vector bosons as well as the vertex
corrections referred to in the Introduction. We have calculated all of
the terms in the non-strange part of the weak Hamiltonian,  to all
orders in perturbation theory using the leading logarithm approximation
of the renormalization group~\cite{dean}.

The result is essentially unchanged for quark-quark scattering, where
the two quarks involved come from the different nucleons. However,
there is another component present, namely when the two quarks involved
come from the same nucleon. At first sight, this appears to be a quark
level description of wavefunction mixing, as was studied in a Regge
model by Soffer and Preparata~\cite{prep}. It was immediately
recognised that their conclusions must be wrong~\cite{bruce}, however,
since they obtained the ANL size and sign for the PVA, but their effect
is the only one that survives to zero energy, thus contradicting the
experimental results at low energy (barring {\bf very} rapidly varying
amplitudes below 10 MeV).

There is a more subtle error in that approach also. It is due to the
fact that the quark-quark scattering in one nucleon can be influenced
by the proximity of the other nucleon in the overall process. To see
this, think of the quarks in a nucleon as wavepackets continually
rescattering on each other due to the attractive strong interaction. A
parity-violating wavefunction mixing occurs when the rescattering of
this pair of quarks (diquark) occurs due to the action of the weak
Hamiltonian rather than the strong. (Note that only the isospin-one,
spin-one, vector diquark can contribute.) However, if there is another
nucleon nearby, a non-negligible probability develops for one of its
quarks to scatter on one of the quarks in the diquark just before the
weak scattering occurs. (Or just after, but the other time-ordering is
intuitively easier to understand.) This injects a four-momentum which
raises the intermediate state diquark (before the weak scattering) to
larger mass scales.  As such, this effect includes all relevant
parity-violating mixing between the nucleon and {\bf higher} mass
baryonic states.  It is {\bf not} constrained by (low energy)
nuclear data on (diagonal) parity violating components of the nucleon
itself.  The probability of the weak scattering has not been otherwise
significantly affected because the `other' quark of the pair was
already `focused' to rescatter with this boosted quark, due to the
strong interactions that shaped their wavefunctions. In nuclear physics
terms, this is a combination of distorted wave (intial and final) and
three-body effects. Aside from the change in the flux relevant to the
scattering process, the rapid growth of the weak scattering strength
with energy means that the parity-violating state mixing can be
markedly increased.

Preston and I did find~\cite{me,us} a PVA growing with energy due to this
quark-diquark scattering effect. Moreover, for reasonable parameter
values, the size of the effect is consistent with the ANL result.
Normalizing to that result and correcting for growth in the strong
cross-section at lower energies, we also predicted~\cite{us} a value
consistent with the smaller, but still positive, PVA found at lower
energy in the LAMPF experiment~\cite{vince}. For completeness,
we should note that in both the quark-quark and the quark-diquark
scattering cases, a single term of the effective weak Hamiltonian
dominates, but its strong interaction enhanced strength is uncertain
by an estimated overall factor of four. This is why we view the energy
dependence of our result as more reliable than the absolute scale.

\section{Response to Criticism}

Simonius and Unger (SU) have taken exception~\cite{su} to our
calculation.  They argue that we must model both the strong and the
weak amplitude for the scattering processes, just as is done in the low
energy meson-baryon descriptions, in order to make a meaningful
prediction of the PVA. In a sense, we agree with them but argue that we
have done a much better job of this than they have.

We take QCD with conventional parameter values to represent the strong
interaction. For the weak interaction, our approximation requires using
a quark-diquark amplitude wherein the diquark undergoes a weak
scattering and one of the quarks in it exchanges a gluon with the
intruding quark. The leading term in the QCD amplitude which overlaps
this, in our approximation, involves a similar structure, with the
internal diquark weak scattering replaced by a gluon exchange to
represent the strong interaction. This last is required because our
diquark representation has been simplified to that of two quarks, each
carrying precisely half of the diquark momentum.  (The diquark momentum
fraction distribution is taken as the complement to the quark momentum
fraction distribution derived experimentally from deep inelastic lepton
scattering.) As a result, without this strong scattering of the two
quarks, there would be an unrealistically poor overlap with the final
state distribution of the two quarks from the weak scattering, which
spreads their strength widely over momentum space. In a sense, we are
modeling the quark wavefunction in the diquark by its perturbative
part. The actual amplitude is stronger, but falls off more rapidly with
momentum transfer until reaching the perturbative level so that, if
anything, we should have an underestimate of the overall size of the
effect.

In addition, however, we use another approximation involving SU(3)
group characters to reduce a twelve $\gamma$-matrix trace to a product
of two traces each involving six $\gamma$-matrices. As a result, only
one of several strong-weak overlap amplitudes survives the color and
Dirac tracings. SU calculate the corresponding strong-strong overlap,
still with only one term, and find a large and rapidly growing (with
energy) contribution to the total strong interaction cross-section.
{}From this their criticism devolves.

However, it is clear on several counts that the SU calculation is
meaningless.  Where we take the measured total cross section for the
strong interaction, SU take only one quark-diquark graph to represent
QCD.  It is straightforward to see that this is not consistent, as
there is no reason to assume that the other QCD graphs vanish.  In
fact, this graph represents neither a complete set nor even a gauge
invariant subset of graphs, and hence, the procedure is not sensible.
This is fortunate, since if their calculation were correct, they would
have shown that the quark-diquark contribution to nucleon-nucleon
scattering proves QCD is inconsistent with data!

That there is some difference is apparent from the fact that SU find
\underline{8 barns} for the nucleon-nucleon total cross section at a
total center-of-mass energy squared ($S$) of 13 GeV$^2$.   If their
calculation were valid, QCD itself would have proven false!  But the SU
calculation cannot be correct since it is well known that the
contribution of such graphs in a renormalizable theory to the total
cross-section cannot grow faster than $ln^{2}(S)$, and their
cross-section grows much faster.  In fact, our initial
results~\cite{me} grow only as $ln(S)$. Furthermore, we improved our
calculation~\cite{us} by including the logarithmic variation of the
strong coupling with momentum transfer scale (which, to carry out the
resulting integrations, required forcing an equality between the
renormalization scale and effective gluon mass, or infrared cutoff).
This causes the PVA to fall asymptotically as $ln(ln(S))/ln(S)$ despite
the nonrenormalizable weak vertex (which, by the way, limits the
applicability of our results to $\sqrt{S} \stackrel{<}{\sim}$ 1 TeV,
where the effect of the W-boson propagator should become apparent), and
so the SU result should fall even faster.  We re-emphasize that their
result does not fall, but rather increases rapidly with $S$.

We are unable to trace the source of the error, since they present only
numerical results.  Note, however, that their problem is reminiscent of
similar ones in QED where gauge invariance has not been properly
implemented.  There, as here, a single graph at a given order can be
larger than the sum, showing that there, as here, arbitrarily picking
out one graph is completely unjustified.  Our effectively single graph
result for the weak PVA numerator came from examining all graphs to
this order, and finding that, {\bf in that particular case}, the
rest were negligible, or vanished.  It is clear this would {\bf not} be
the case for the QCD denominator.

Finally, we note that SU implicitly propose summation of the large
coupling constant, divergent perturbation series for QCD. No such
calculation would be credible. Even the operator-product type of
analysis for summing leading logarithms to calculate strong interaction
enhancements to the weak amplitude is subject to serious criticism,
although its employment is standardly accepted. Rather, it is our
experience that the {\bf leading} term in a QCD calculation, with normal
parameter values, gives a good representation of the physics in any
given process, up to an overall strength factor. This is why we used
the experimental value for the total strong-interaction cross-section,
checking only that the {\bf leading} graphs for this are consistent the
experimental results for the (normally accepted) parameter values that
we used in the rest of the calculation of the PVA.

\section{A Proposal for an `Improved' Calculation}

Our calculation can, of course, be improved. We have recently realized
that, at the cost of going to a level of complexity involving traces of
products of eight $\gamma$-matrices, we can redo the quark-diquark
scattering calculation in a different way. It requires modeling the
diquark wavefunction, and we were initially reluctant to add such an
ans\"{a}tz. It will, however, allow us to calculate to one lower order
in the strong interaction, since we can now use this wavefunction to
provide for the overlap with the weak scattering amplitude.  Thus, the
strong interaction will appear only in the same order in the numerator
and (conceptually) in the denominator of our PVA estimate. As we just
argued in the previous section, a lower order calculation in QCD should
be more reliable (at least, in that regard). Furthermore, such a
calculation will provide a qualitative check on our result. From the
structure of the integrals, we can see no reason for a markedly
different result.

\section{Effect of the QCD Spin Crisis}

Our model is based on the heretofore conventional wisdom that all of
the nucleon spin is carried by the valence quarks~\cite{crisis}. If
the sea and gluons are highly polarized, then graphs for the weak
amplitude which we have ignored, such as polarized gluon-gluon
scattering,  could become important.  We would find this hard to credit
except for one consideration:  the two-phase vacuum model of
confinement involves chromo-electric and -magnetic fields.  These could
carry significant spin, polarizing the sea quarks to produce a precise
cancellation for an ``empty'' perturbative vacuum bubble.  Introduction
of polarized valence quarks would certainly disturb this cancellation,
and it is precisely at small Bjorken~$x$ where one would expect the
largest effect.  We speculate that this is related to high-p$_T$
polarization phenomena~\cite{hipt}, becoming significant  when the
p$_T$ is large enough that the hard scattering involved occurred in
{\bf one} polarization region.  However, this speculation and the
effect of these considerations on the PVA require considerable
additional effort before any conclusions can be drawn.  Conversely, the
measured high energy PVA may be an important constraint for
interpreting the results of such a theoretical study.

The situation may not be of too great concern, however, since our PVA
contibution comes dominantly from harder scattering of quarks at higher
Bjorken~$x$. As Kunz, Mulders and Pollock have shown~\cite{kmp}, the
valence quark spin distribution is not significantly different from
expectation. Indeed, they find that the entire measured result is not
unreasonable, since any low p$_T$ quark model will evolve spin from the
valence constituent quarks to the sea and gluons as Q$^2$ increases.
Therefore, we do not expect this interesting development to invalidate
our calculation or conclusions.

\section{Conclusion}

In summary, we have presented a simple model which, as for deep
inelastic structure functions or Drell-Yan lepton-pair production,
cannot supply an accurate prediction of the PVA at a given energy, but
which should be valid for the (strong) energy dependence of the PVA at
high energies.  Due to approximations made in evaluating it, the model
is not applicable above 1 TeV.  Amazingly, it is consistent with data
between 6 and 1.5 GeV/c, when the variation of the total
nucleon-nucleon cross section between those beam momenta is taken
crudely into account.

We predict that an experiment at Brookhaven should expect to find a PVA
$\sim$ 10$^{-5}$ and one at Fermilab, almost 10$^{-4}$.  Naturally, the
prudent experimenter will design for an order of magnitude better
sensitivity than these predictions, if possible. More importantly,
perhaps, we have shown that, if PVA's at these levels are not observed,
one may interpret such a result as evidence for the existence of new,
right-chiral weak interactions. Calculations are in progress to provide
a firm numerical link between the value of (or upper bound on) the PVA
at a given energy and the upper bound on the mass scale for the new
weak vector boson. A confirmation of the PVA at somewhat lower energies
is also needed.

\section{Acknowledgments}

The work reported here was done in collaboration with G.\  J.\
Stephenson, Jr.\ , Dean Preston and Kim Maltman. It was supported in
part by the US DOE.


\end{document}